\documentclass[runningheads]{llncs}
\usepackage[T1]{fontenc}

\usepackage{graphicx}
\usepackage{booktabs}
\usepackage{amsmath}
\usepackage{amssymb} 
\usepackage{microtype}\emergencystretch=1em

\begin{document}
\title{MultiHedge: Adaptive Coordination\\via Retrieval-Augmented Control}
\titlerunning{MultiHedge: Adaptive Coordination via RAG}

\author{Feliks Bańka\orcidID{0009-0005-1973-5861}\\\and
Jarosław A. Chudziak\orcidID{0000-0003-4534-8652}}
\authorrunning{F. Bańka and J. A. Chudziak}

\institute{The Faculty of Electronics and Information Technology, \\ 
Warsaw University of Technology, Warsaw, Poland \\
\email{\{feliks.banka.stud, jaroslaw.chudziak\}@pw.edu.pl}}
\maketitle              

\begin{abstract}

Decision-making under changing conditions remains a fundamental challenge in many real-world systems. Existing approaches often fail to generalize across shifting regimes and exhibit unstable behavior under uncertainty. This raises the research question: can retrieval-augmented LLM coordination improve the robustness of modular decision pipelines? We propose MultiHedge\footnote{Code and experimental configuration files are publicly available at: \url{https://github.com/latent-systems-lab/MultiHedge}}, a hybrid architecture where an LLM produces structured allocation decisions conditioned on retrieved historical precedents, and execution is grounded in canonical option strategies. In a controlled evaluation using U.S. equities, we compare MultiHedge to rule-based and learning-based baselines. The key result is that memory-augmented retrieval confers greater robustness and stability than increasing model scale alone. Our paper contributes a controlled computational study showing that memory and architectural design play a central role in robustness in modular decision systems.

\end{abstract}

\section{Introduction}

Financial markets are characterized by frequent regime shifts and volatility bursts, complicating risk management and making risk management unstable~\cite{Shu2024}. Even established option strategies such as collars and straddles provide only partial protection if not coordinated over time~\cite{BankaAciids}. Selecting and adjusting these hedges under changing market conditions is a sequential control problem involving both discrete and continuous decisions~\cite{Puterman1994}.

Recent advances in large language models (LLMs) have enabled structured workflows that integrate heterogeneous signals and produce auditable rationales~\cite{kirtac2024llm}. Retrieval-based memory mechanisms offer a lightweight way to condition decisions on similar past states, improving robustness under distribution shift~\cite{Aamodt1994}. However, most hedging frameworks focus on end-to-end learning or static overlays, rarely studying retrieval-conditioned coordination over modular, interpretable decision primitives~\cite{malekzadeh2024exdrlhedgingheavylosses}.

This raises a central research question: can retrieval-augmented coordination enhance the robustness of modular decision pipelines under distributional shift and regime instability? We address this by introducing \textit{MultiHedge} (Section~\ref{sec:overview}), a hybrid computational architecture in which the LLM serves as a bounded heuristic approximator within a constrained optimization loop. Allocation decisions are strictly conditioned on retrieved historical analogues, while execution is handled by formally defined option-hedging modules. This design dynamically coordinates canonical strategies, separating stochastic reasoning from symbolic execution to align language-model inference with constrained portfolio control~\cite{Lewis2020}.

The main contributions of this work are as follows. First, we formulate retrieval-conditioned coordination as a structured sequential decision problem. Second, we introduce a modular hybrid architecture that separates reasoning, action generation, and execution within a reproducible workflow. Third, we present a controlled computational study—including limited robustness checks—that isolates the role of memory and coordination in improving robustness beyond model scale.
\section{Related work}

Large language models (LLMs) and retrieval-augmented generation (RAG) workflows have been integrated into structured decision systems to ground outputs in external evidence and improve interpretability~\cite{Guo2024,Lewis2020}. In finance, such systems support sentiment extraction, event interpretation, and tool-augmented trading components~\cite{finagent}. Retrieved episodes provide precedents that align model reasoning with historical cases~\cite{Aamodt1994}. Conditioning decisions on retrieved experiences improves robustness under distribution shift and non-stationarity~\cite{xiao2025}, with episodic memory acting as a non-parametric prior that grounds inference in observed trajectories rather than purely parametric knowledge. In structured decision architectures, LLMs increasingly operate as bounded reasoning modules over constrained action spaces, where outputs are parsed and executed by deterministic components~\cite{Guo2024,Macal2010}. In \textit{MultiHedge}, episodic recall informs allocation decisions over option-level primitives, coupling regime inference with historical outcome alignment~\cite{finagent}.

From a computational-science perspective, modular control architectures are standard mechanisms for robustness under non-stationarity and partial observability~\cite{Puterman1994,Liberzon2003}. Specialist modules capture complementary behaviours, while a coordinator performs context-dependent selection under explicit constraints. Learning-based hedging and risk-control frameworks demonstrate that such policies can operate under frictions when embedded into a well-defined execution model~\cite{buehler2019deep}, and classical parameterised primitives remain widely used due to predictable cost--benefit trade-offs~\cite{Macal2010}. Existing financial LLM systems such as FinAgent primarily emphasize signal interpretation and tool use~\cite{finagent}. Ensemble and mixing approaches improve downside stability~\cite{banka2025deltahedgepacis,Szydlowski2024b}, yet they rarely formalise retrieval-conditioned coordination over fixed interpretable primitives within a constrained sequential control and backtesting framework. Our work addresses this gap by explicitly separating retrieval, allocation synthesis, and symbolic execution in a reproducible computational workflow.

\section{MultiHedge: Hybrid Architecture, Episodic Memory, and Modular Coordination}

\begin{figure}[t!]
	\centering
	\includegraphics[width=0.85\linewidth]{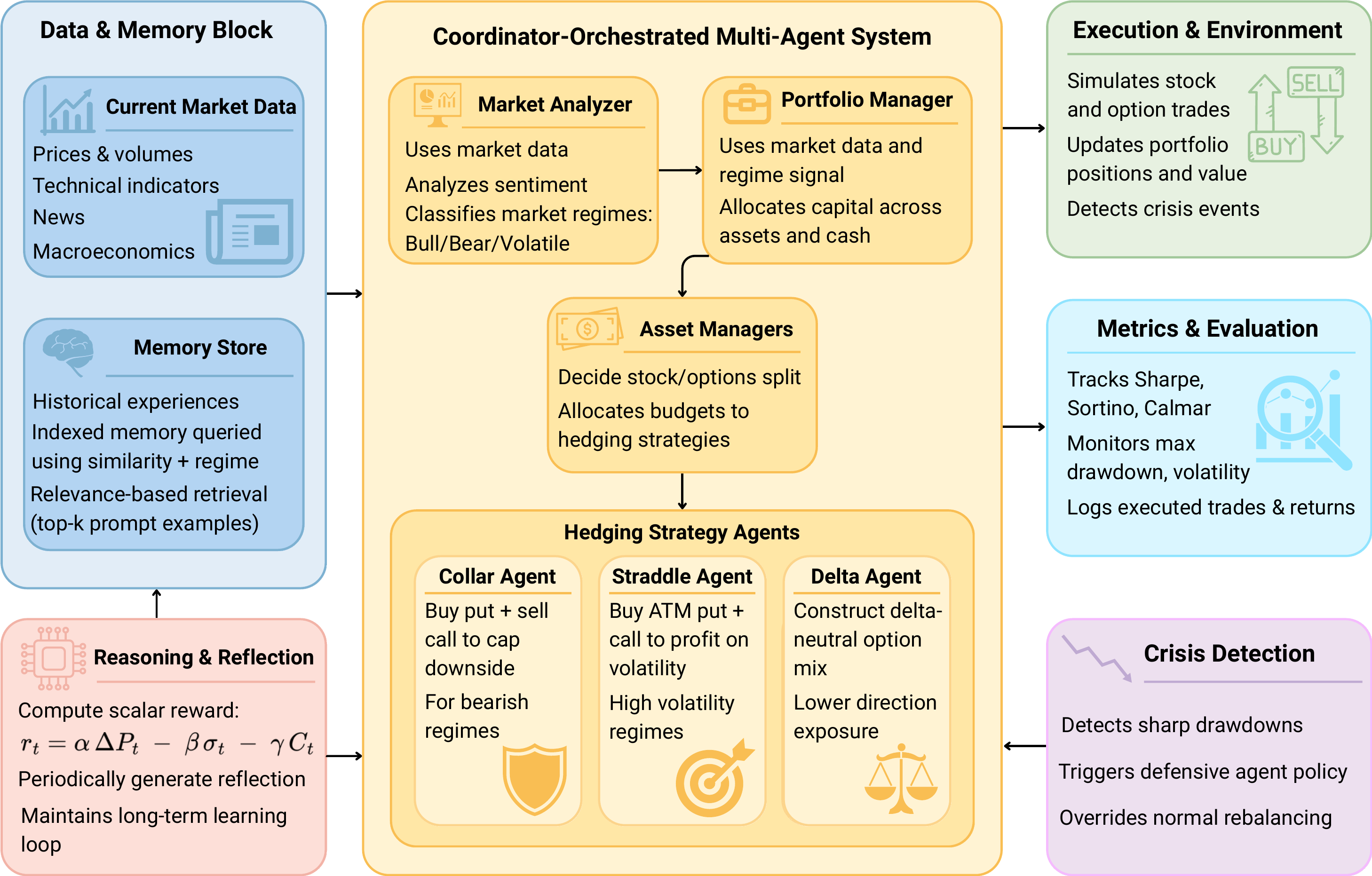}
	\caption{MultiHedge system architecture: modular components, episodic memory, and hybrid control pipeline.}
	\label{fig:architecture}
\end{figure}

This section details the MultiHedge architecture, which integrates episodic memory, modular coordination, and a hybrid control pipeline. Figure~\ref{fig:architecture} illustrates the system's main components and their interactions.

\label{sec:framework}
We formulate dynamic hedging as a constrained sequential control problem under partial observability, where portfolio allocation must remain bounded under regime shifts and transaction frictions. Partial observability arises because regime and volatility states are latent.

Formally, the problem is represented as a Partially Observable Markov Decision Process (POMDP)~\cite{Puterman1994}:
$$\langle \mathcal{S}, \mathcal{A}, \mathcal{O}, \mathcal{T}, \mathcal{R} \rangle,$$
where $\mathcal{S}$ is the latent market--portfolio state space, $\mathcal{A}$ is a mixed discrete--continuous action space, $\mathcal{O}$ are observable signals (prices, volatility, news features), $\mathcal{T}$ denotes transition dynamics, and $\mathcal{R}$ is a risk-adjusted reward functional. \textit{MultiHedge} implements a hybrid computational architecture that combines symbolic option-hedging structures, retrieval-augmented memory, and language-model-based reasoning within this decision process.

\subsection{System Overview}
\label{sec:overview}

As shown in Figure~\ref{fig:architecture}, MultiHedge consists of several interacting modules: an LLM-based inference layer, a portfolio and allocation layer, symbolic hedging agents, and a deterministic safety layer. The figure provides a visual summary of how these components coordinate to produce memory-augmented decisions.

The LLM processes multimodal signals and outputs regime classification and structured allocation candidates. The portfolio layer implements a risk-adjusted objective, while hedging agents execute canonical option strategies. The safety layer enforces stability constraints by overriding actions during drawdown events, ensuring bounded risk and robust system behavior.

\subsection{Decision Process and Action Formulation}
\label{sec:decision-flow}

We model hedging as a constrained decision process where, at each trading day $t$, the system observes a partially informative state and produces structured allocation and hedging actions subject to feasibility and stability constraints.

Formally:
\begin{align}
w_t &= \pi_{\text{alloc}}(s_t, R_t, E_t), \\
a_{i,t}^H &= \pi_{\text{agent}}^H(s_t, w_t), \quad \forall H \in \{\text{collar, straddle, delta-neutral}\}.
\end{align}

Under this formulation, the LLM acts as a constrained reasoning engine. It produces structured outputs that are grounded via retrieval and executed through interpretable primitives. These outputs are parsed and validated against explicit financial constraints, including budget, liquidity, and risk limits. This ensures that all actions remain feasible, interpretable, and bounded by the system's operational parameters at every step.

To mitigate hallucinations and enforce stability, a deterministic safety layer overrides the LLM's outputs when drawdown thresholds are breached. Deterministic inference (temperature zero) further reduces stochasticity, ensuring predictable behavior under stress.

\noindent
The retrieval-augmented policy update is given by:
\begin{align*}
	\pi^* \leftarrow \arg\max_{\pi} \; \mathbb{E}_{(s, a) \sim \mathcal{M}} \left[ \mathrm{sim}(\phi(s, a), \phi(s', a')) - \lambda \, \mathcal{L}(\pi) \right]
\end{align*}
where $\mathrm{sim}(\cdot,\cdot)$ denotes embedding similarity, $\phi$ is the embedding function, $\lambda$ controls retrieval strength, and $\mathcal{L}(\pi)$ is a regularization term. Under this formulation, retrieval constrains policy updates toward historically aligned trajectories, acting as a non-parametric regularizer under distributional shift.

\subsection{Memory-Augmented Reasoning via Metric Retrieval}
\label{sec:memory}

\begin{figure}[t!]
\centering
\includegraphics[width=0.99\linewidth]{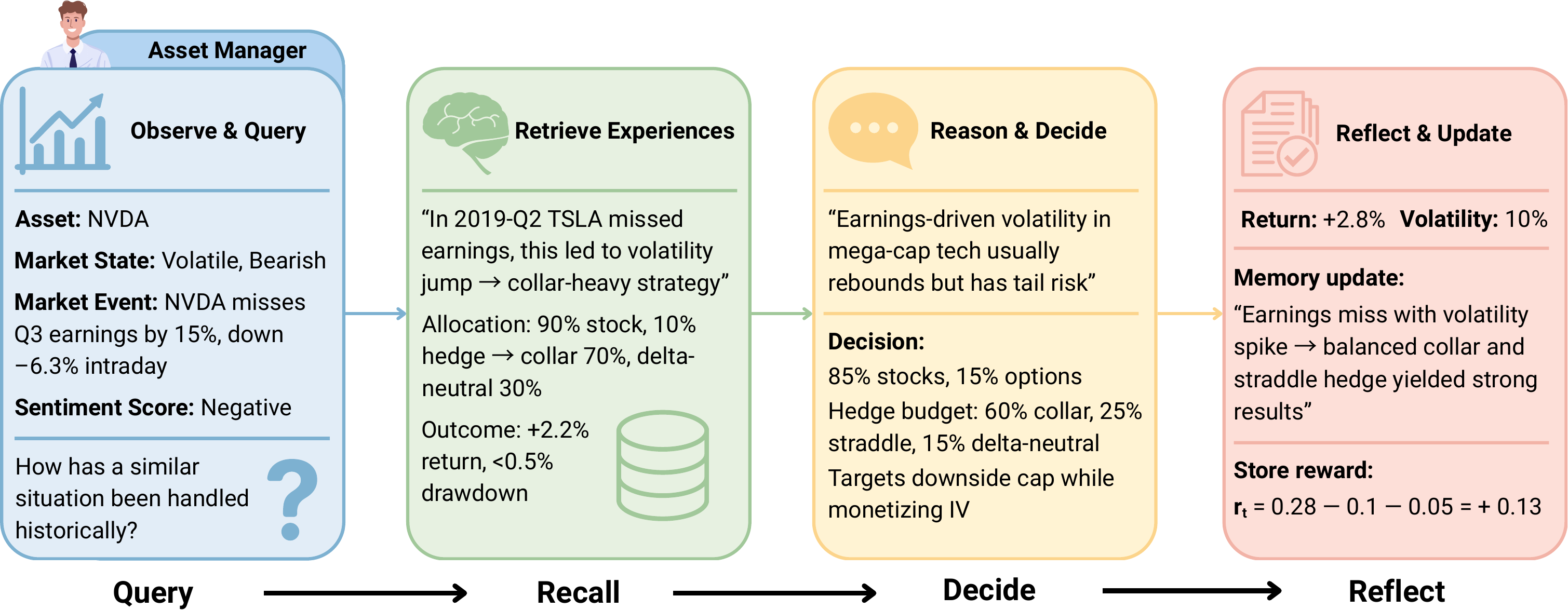}
\caption{Episodic memory loop in MultiHedge: past state–action–outcome tuples are retrieved via similarity search, used to condition current allocation decisions, and then stored for future reuse.}
\label{fig:prompts}
\end{figure}

As illustrated in Figure~\ref{fig:prompts}, the system follows a retrieval–reason–act–update loop driven by a non-parametric episodic memory buffer $\mathcal{M}$. Past decision episodes are stored within this buffer as state–action–outcome tuples, establishing a continuous cycle where historical trajectories are retrieved to condition current actions and then updated for future reuse.

During the decision phase, the controller retrieves the top-$k$ most similar episodes using cosine similarity and anchors its allocation logic to their realized outcomes. By explicitly conditioning on these retrieved precedents, the system reuses empirically validated strategies in analogous contexts rather than relying solely on parametric generalization.
\section{Experimental Evaluation under Changing Market Conditions}

We evaluate MultiHedge in a setting where market conditions vary over time. The system is compared against classical and learning-based baselines, and we perform targeted ablations to isolate the effects of memory, coordination, and model scale.

\subsection{Experimental Setup and Robustness Metrics}

We use daily historical market data for large-cap equities (AAPL, TSLA, NVDA) from January 2021 to December 2023, with memory index calibration on 2016–2020. Trading costs and liquidity constraints are enforced, and the LLM controller GPT-4.1 operates deterministically ($T=0.0$).

Evaluation relies on core performance and tail-risk metrics: Total Return (TR), Sharpe Ratio (SR; return per unit of total volatility), Maximum Drawdown (MD; largest peak-to-trough drop), and Conditional Value at Risk (CVaR). We additionally track auxiliary stability indicators: Sortino (SoR) and Calmar (CaR) ratios for downside-adjusted returns, Value at Risk (VaR), worst-month return (WM), downside deviation (DDv; volatility of negative returns), time spent in severe drawdown (TDD $>10\%$), and maximum consecutive loss days (MCL). These metrics exhibit consistent trends and reinforce the observed improvements in risk-adjusted performance. Auxiliary robustness checks, including broad-market stress and synthetic regime-switching tests, further confirm the stability.

\begin{table*}[t!]
  \centering
  \caption{Performance Comparison: Core Metrics and Downside Risk}
  \label{tab:comparison_benchmark}
  \vspace{1mm}
  \begin{tabular*}{\textwidth}{@{\extracolsep{\fill}}lcccccc}
    \toprule
    \textbf{Strategy} & \textbf{SR} & \textbf{TR (\%)} & \textbf{MD (\%)} & \textbf{CVaR (\%)} & \textbf{WM (\%)} & \textbf{DDv (\%)} \\
    \midrule
    Buy \& Hold   & 0.81 & 107.08 & 53.08 & 5.36 & -21.96 & 25.34 \\
    Equal Weight  & 0.77 & 95.34  & 53.53 & 5.41 & -21.29 & 25.73 \\
    \midrule
    PPO           & 0.74 & 66.46  & 39.72 & 3.95 & -15.59 & 18.68 \\
    FinAgent      & 0.63 & 69.74  & 54.60 & 5.99 & -21.92 & 28.69 \\
    FinRL         & 0.56 & 57.08  & 61.51 & 6.48 & -25.19 & 30.78 \\
    \midrule
    \textbf{MultiHedge} & \textbf{1.69} & \textbf{152.08} & \textbf{16.22} & \textbf{2.29} & \textbf{-4.58} & \textbf{10.93} \\
    \bottomrule
  \end{tabular*}
\end{table*}

\subsection{Benchmark Results}
Table~\ref{tab:comparison_benchmark} shows that MultiHedge consistently improves both performance and risk characteristics relative to all baselines. The Sharpe ratio increases substantially compared to classical strategies, while total return improves without a corresponding increase in risk.
The most pronounced gains appear in downside risk. Maximum drawdown is reduced by over 70\% relative to Buy \& Hold (53.08\% $\rightarrow$ 16.22\%), and CVaR is more than halved. The worst-month loss improves from approximately $-22\%$ to $-4.6\%$, indicating strong mitigation of extreme negative events. Similarly, downside deviation is reduced by more than 50\%, suggesting that improvements are not limited to isolated tail events but reflect a broader compression of downside volatility.

These effects indicate that MultiHedge not only improves average performance but systematically reshapes the loss distribution, reducing both the magnitude and persistence of adverse regimes. Consistent improvements across additional monitored metrics (SoR, CaR, VaR, TDD, MCL) further support this interpretation. Auxiliary tests under stress and regime-switching scenarios confirm that these gains remain stable across perturbations, validating retrieval-conditioned coordination as a robust mechanism for non-stationary environments.

\subsection{Architectural Ablation Studies}
\label{subsec:ablation}

We conduct ablation studies to isolate the effects of memory, stochasticity, and model scale. GPT-4o Mini is used as a reduced-capacity baseline to assess the effect of model scale under fixed architecture.
Table~\ref{tab:llm_ablation} shows that memory is the dominant factor driving robustness. Removing memory leads to a near tripling of drawdown (16.22\% $\rightarrow$ 46.18\%) and a substantial deterioration in tail risk (CVaR: 2.29 $\rightarrow$ 4.78). The degradation is also visible in worst-month losses and downside deviation, indicating both more frequent and more severe adverse outcomes.

Introducing stochasticity further degrades performance, suggesting that deterministic, constrained inference is important for stability in sequential decision settings. In contrast, reducing model capacity results in a comparatively smaller performance drop across all metrics.
While we consider a limited comparison of model scales, these results indicate that architectural components—particularly memory—have a stronger impact on robustness than model size in this setting.
\begin{table}[t!]
\centering
\caption{Ablation of Control Architecture: Core and Downside Risk Metrics}
\label{tab:llm_ablation}
\begin{tabular}{lcccccc}
\toprule
\textbf{Configuration} & \textbf{SR} & \textbf{TR (\%)} & \textbf{MD (\%)} & \textbf{CVaR (\%)} & \textbf{WM (\%)} & \textbf{DDv (\%)} \\
\midrule
MultiHedge (GPT-4.1)  & 1.69 & 152.08 & 16.22 & 2.29 & -4.58 & 10.93 \\
GPT-4o Mini           & 1.14 & 97.44  & 20.41 & 2.87 & -9.26 & 13.77 \\
Stochastic ($T=0.7$)  & 0.96 & 84.58  & 28.28 & 3.28 & -12.27 & 15.86 \\
No Memory             & 0.80 & 93.94  & 46.18 & 4.78 & -19.95 & 22.89 \\
\bottomrule
\end{tabular}
\end{table}

\section{Discussion and Future Work}
The empirical outperformance of MultiHedge over reinforcement learning baselines highlights a broader shift in designing computational financial agents. Deep RL approaches, such as PPO or FinRL~\cite{Liu2022}, often struggle with sudden regime shifts because they optimize over historically stationary reward landscapes~\cite{buehler2019deep}. Conversely, purely LLM-based trading systems~\cite{finagent} risk unstable, unconstrained execution. By restricting the LLM to a coordinating role and offloading execution to deterministic primitives, our architecture achieves robust downside protection without requiring end-to-end retraining. This suggests that grounding modular components via episodic memory can provide a more stable path to robustness than relying solely on parametric scale~\cite{Lewis2020,Guo2024}.

Despite these advantages, limitations remain. Restricting evaluation to large-cap equities limits generalisation to asset classes with different microstructures. Furthermore, while deterministic inference reduces instability, the system remains sensitive to retrieval quality, prompt specification, and inherent LLM training biases or data leakage.

Future work will explore three directions: incorporating term-structure and volatility surface features to improve regime sensitivity; integrating risk-sensitive objectives to stabilise extreme-event performance~\cite{malekzadeh2024exdrlhedgingheavylosses}; and generalising the architecture to structured multi-agent settings where specialised controllers interact via explicit coordination protocols~\cite{Macal2010}.
\section{Conclusion}
\label{sec:conclusion}
Financial decision-making under market regime shifts often suffers from instability and increased downside risk. While recent approaches leverage large language models or end-to-end learning, it remains unclear whether their success stems from raw model scale or structural design. This raises a key question: can conditioning decisions on retrieved historical precedents systematically improve system robustness?

To address this, we introduced \emph{MultiHedge}, a hybrid architecture that combines structured decision rules with LLM-based reasoning guided by retrieved historical examples. Our results show consistent improvements in performance and substantial reductions in downside risk compared to both rule-based and learning-based baselines. Ablation studies indicate that memory and system design play a larger role than model size in driving these gains. Overall, this paper contributes to the field of robust decision systems by showing that combining memory with modular decision structures can improve performance under changing conditions.

\bibliographystyle{splncs04}
\bibliography{main/mybibfile}

\end{document}